\documentclass[runningheads]{llncs}
\usepackage[T1]{fontenc}
\usepackage{graphicx}
\usepackage{url}
\usepackage{hyperref}
\usepackage{amsmath}
\usepackage{amssymb}
\usepackage{dsfont}
\usepackage{booktabs}
\usepackage{multirow}
\usepackage{rotating}
\usepackage{tablefootnote}
\usepackage{cite}

\usepackage{microtype}
\begin{document}
\title{FNH-TTS: Mixture-of-Experts Duration Modeling for Robust Neural Speech Synthesis}
\titlerunning{FNH-TTS}
\iffalse
\author{First Author\inst{1}\orcidID{0000-1111-2222-3333} \and
Second Author\inst{2,3}\orcidID{1111-2222-3333-4444} \and
Third Author\inst{3}\orcidID{2222--3333-4444-5555}}
%
\authorrunning{F. Author et al.}
\institute{Princeton University, Princeton NJ 08544, USA \and
Springer Heidelberg, Tiergartenstr. 17, 69121 Heidelberg, Germany
\email{lncs@springer.com}\\
\url{http://www.springer.com/gp/computer-science/lncs} \and
ABC Institute, Rupert-Karls-University Heidelberg, Heidelberg, Germany\\
\email{\{abc,lncs\}@uni-heidelberg.de}}
\fi
\iffalse
\author{Anonymous Authors}
\institute{Anonymous Institution\\
\email{anonymous@example.com}}
\fi

\author{Qingliang Meng\inst{1} \and 
Yuqing Deng\inst{1} \and
Wei Liang\inst{1} \and
Limei Yu\inst{1} \and
Huizhi Liang\inst{2}\and
Tian Li\inst{2}
}
\authorrunning{Q. Meng et al.} % abbreviated author list (for running head)

\institute{Megatronix, Beijing, China \and
Newcastle University, Newcastle Upon Tyne, United Kingdom\\
\email{mengqingliang9485@163.com}, \email{yuqing.deng.i@megatronix.com}, \email{wei.liang@megatronix.com}, \email{limei.yu@megatronix.com}, \email{huizhi.liang@newcastle.ac.uk}, \email{t.li56@newcastle.ac.uk}}

\maketitle              % typeset the header of the contribution
\begin{abstract}

Natural and human-like speech depends on the coordination between prosodic timing and acoustic realization: duration modeling shapes rhythmic structure, while waveform generation determines whether that structure is rendered naturally. In natural speech, duration patterns vary across linguistic contexts and speakers, requiring a TTS system both to capture this variability and to faithfully realize it in the waveform. To address these challenges, we propose FNH-TTS, a VITS-based end-to-end system that jointly improves duration modeling and waveform generation. A mixture-of-experts duration predictor (MoE-DP) uses multiple experts and routing jointly conditioned on linguistic context and speaker information to model diverse duration patterns.
For waveform generation, we adopt an inverse short-time Fourier transform (ISTFT)-based generator, providing a more direct and efficient synthesis path. We further employ multi-resolution and sub-band discriminators for fine-grained temporal and spectral adversarial supervision, thereby supporting natural waveform synthesis.
Experiments on LJSpeech, VCTK, and LibriTTS show that FNH-TTS achieves the highest mean MOS on LJSpeech and VCTK and the highest duration-category accuracy on LibriTTS among the compared systems, together with competitive waveform reconstruction and substantially faster vocoder inference. Controlled analyses further show that MoE-DP primarily drives the duration-modeling gains, while the vocoder-side components make complementary contributions to synthesis quality and efficiency.

\end{abstract}

\section{Introduction}

Text-to-speech (TTS) aims to synthesize natural speech from text in a desired speaking style. Prosodic timing plays a central role in perceived naturalness because it governs how speech unfolds over time. Phoneme durations vary with linguistic context and speaker-specific speaking habits, producing diverse patterns of rhythm, emphasis, and speaking rate. Natural synthesis therefore requires a system to capture these duration patterns and render the resulting acoustic transitions coherently in the waveform.

Autoregressive (AR)~\cite{wang2017tacotron,li2019neural,chen2024vall} and non-autoregressive (NAR)~\cite{ren2019fastspeech,kim2020glow,ju2024naturalspeech} TTS systems model this temporal organization differently. AR systems determine duration implicitly during sequential generation, whereas NAR systems typically predict duration explicitly to construct text-to-speech alignment. In many NAR systems, this role is assigned to a dedicated Duration Predictor (DP). Capturing diverse contextual and speaker-dependent duration patterns within such an explicit predictor remains challenging~\cite{mehta2024should}.

Recent studies have analyzed and improved duration modeling from different perspectives. StyleTTS~\cite{li2025styletts} points out that duration labels derived from hard alignment, soft alignment, or Monotonic Alignment Search (MAS)~\cite{kim2020glow} are not true ground-truth labels, and overfitting to such labels can reduce speech naturalness. Mehta et al.~\cite{mehta2024should} further highlight the importance of duration modeling and call for more direct methods to assess prosodic complexity. VITS2~\cite{kong2023vits2} improves the original VITS~\cite{kim2021conditional} framework through adversarial learning to enhance duration prediction, particularly in multi-speaker scenarios. Nevertheless, these approaches mainly focus on alignment quality, supervision strategy, or adversarial objectives, while the internal structure of the DP itself remains relatively underexplored for capturing diverse duration patterns.

Improving duration prediction addresses the temporal organization of speech, but these temporal patterns must ultimately be realized by the waveform generator. VITS adopts a HiFi-GAN-style generator that synthesizes waveforms through learned temporal upsampling, which previous studies have associated with tonal artifacts and spectral distortions~\cite{kawamura2023lightweight,siuzdak2023vocos,bak2023avocodo}. The accompanying adversarial discriminators primarily examine waveform structures at different scales and periods, whereas frequency-band-specific distortions receive less explicit attention~\cite{kong2020hifi,bak2023avocodo}. These limitations are relevant to our goal: representing more diverse duration patterns requires the waveform stage to render acoustic events with different temporal spans while preserving continuity and spectral detail. This motivates a waveform-generation design with a more direct synthesis path and finer temporal and spectral supervision.

In this paper, we propose FNH-TTS, a VITS-based NAR TTS system that jointly improves duration modeling and waveform generation. 
For duration modeling, we introduce a Mixture-of-Experts Duration Predictor (MoE-DP). Its router jointly uses contextual phoneme representations and speaker embeddings to assign phoneme-level states to different experts, providing additional capacity for modeling contextual and speaker-dependent duration patterns.
For waveform generation, we adapt the ISTFT-based architecture of VOCOS~\cite{siuzdak2023vocos}, originally developed as a separate neural vocoder, to operate on the VITS posterior latent representation and train it jointly with the complete TTS system. We further incorporate the Collaborative Multi-Band Discriminator (CoMBD) and Sub-Band Discriminator (SBD)~\cite{bak2023avocodo} to provide complementary supervision across temporal resolutions and frequency bands. Together, MoE-DP models diverse duration patterns, while the generator and discriminators support their efficient and natural acoustic realization.

Our key contributions are as follows:
\begin{itemize}
    \item We introduce a Mixture-of-Experts Duration Predictor (MoE-DP), which uses multiple experts and context- and speaker-conditioned routing to model diverse duration patterns. Duration-specific evaluations and controlled analyses demonstrate its contribution to duration modeling.
    
    \item We develop FNH-TTS by combining MoE-DP with a VOCOS-style generator and CoMBD/SBD in an end-to-end VITS framework. The generator supports efficient frequency-domain synthesis, while CoMBD and SBD provide complementary multi-resolution and sub-band supervision for waveform quality.
    
    \item We conduct leave-one-component-out ablations and targeted duration and vocoder evaluations to distinguish the roles of the proposed components. The results further show that WER reflects content accuracy and intelligibility and should be interpreted alongside other metrics when assessing overall synthesis quality.
\end{itemize}

\section{Methodology}

\begin{figure*}[htbp]
\vskip -20pt
  \centering
  \includegraphics[width=\linewidth]{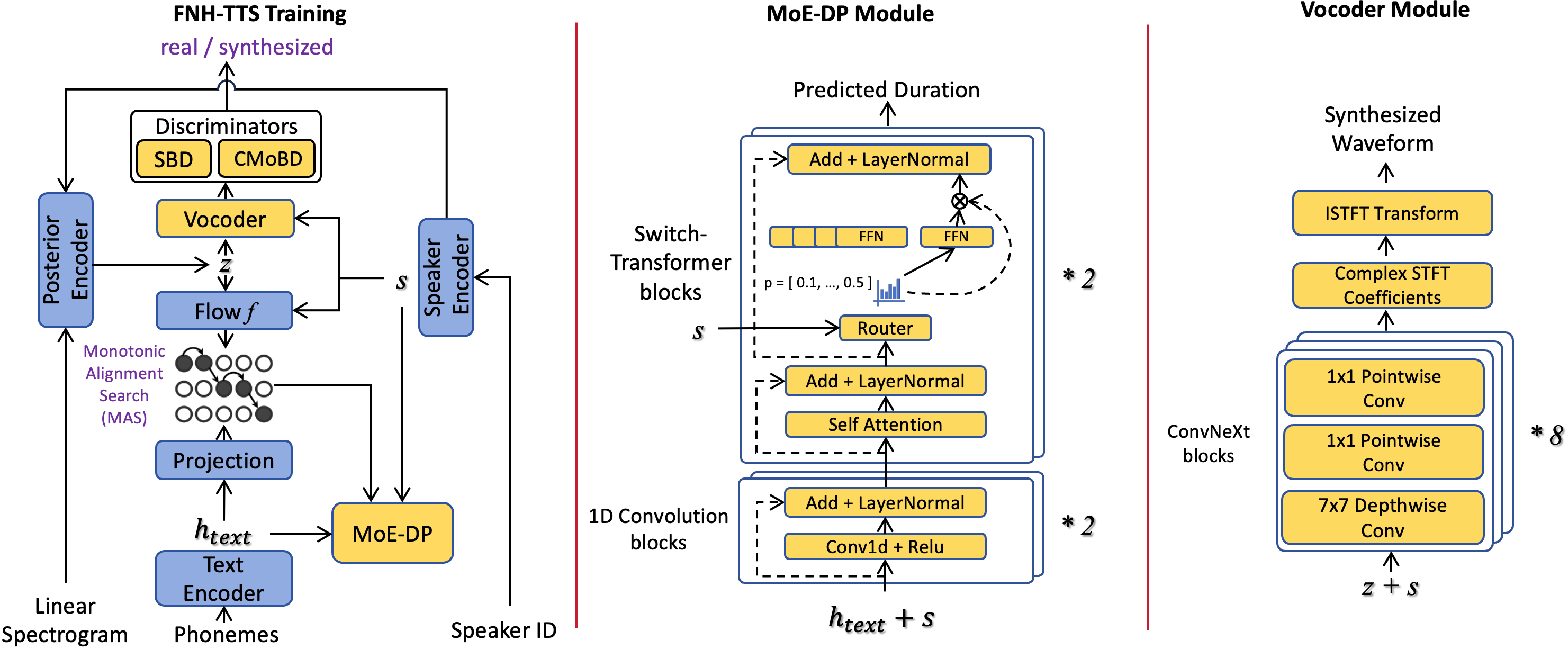}
  \vskip -10pt
  \caption{The overall framework of the FNH-TTS system. The blue modules represent components inherited from the original VITS, while the yellow modules indicate newly introduced modifications.}
  \label{fig:Overall}
  \vskip -15pt
\end{figure*}

Our system, FNH-TTS, is built upon VITS and retains its Text Encoder, Speaker Encoder, Posterior Encoder, and Flow modules. We mainly modify the Duration Predictor (DP) and vocoder-related components, as shown in Fig.~\ref{fig:Overall}. The overall optimization objective is formulated as follows:
\vskip -15pt
\begin{align}
\mathcal{L} = \mathcal{L}_{rec} + \mathcal{L}_{kl} + \mathcal{L}_{dur} + \mathcal{L}_{adv} + \mathcal{L}_{gen}.
\label{eq:allLoss}
\end{align}
\vskip -5pt

$\mathcal{L}_{rec}$ and $\mathcal{L}_{kl}$ correspond to the reconstruction and KL losses used by the original VITS modules, including the Text Encoder, Speaker Encoder, Posterior Encoder, and Flow. These modules and losses are retained without modification.

$\mathcal{L}_{dur}$ supervises the DP. We replace the original DP with the proposed Mixture-of-Experts Duration Predictor (MoE-DP). This module belongs to the category of Deterministic Duration Predictors (DDP)~\cite{mehta2024should}, as it directly predicts phoneme durations. $\mathcal{L}_{dur}$ consists of two components: $\mathcal{L}_{mas}$, computed from the Monotonic Alignment Search (MAS)~\cite{kim2020glow} result, and $\mathcal{L}_{aux}$, an auxiliary loss for balancing the workload among experts. Details are provided in Section~\ref{sec:MoE-DP}.

$\mathcal{L}_{adv}$ and $\mathcal{L}_{gen}$ denote the discriminator-side adversarial loss and the generator-side vocoder loss, respectively. Following the joint design of duration modeling and waveform generation, we integrate a VOCOS-style generator with a Collaborative Multi-Band Discriminator and a Sub-Band Discriminator. The generator supports efficient ISTFT-based waveform synthesis, while the discriminators provide complementary temporal and spectral adversarial supervision. Details are provided in Section~\ref{sec:vocoder}.

\subsection{MoE Duration Predictor}
\label{sec:MoE-DP}

The Mixture-of-Experts (MoE) routing mechanism, initially introduced by~\cite{shazeer2017sparsely}, increases model capacity through sparse expert routing without substantially increasing inference costs. A typical MoE layer consists of a router and a set of experts. 
The router dynamically assigns each token to suitable experts, allowing different experts to process different input patterns.
Switch Transformer~\cite{fedus2022switch} further applies MoE to the feed-forward network (FFN) of Transformer blocks, enabling efficient capacity scaling.

Building on this idea, we propose a DP architecture composed of 1D convolution blocks and Switch-Transformer blocks, as shown in Fig.~\ref{fig:Overall}. The input to the MoE-DP is $h_{text}+s$, where $h_{text}$ denotes the hidden representation encoded from the phoneme sequence by the Text Encoder, and $s$ denotes the speaker embedding. The router jointly uses the contextual representation and speaker embedding $s$, enabling the routing decisions to reflect both linguistic context and speaker-specific information. The routing and expert processing mechanism is formulated as follows:
\vskip -10pt
\begin{align}
	y = \sum_{i\in\Psi}p_{i}(x + s)E_{i}(x),
\label{eq:moe}
\end{align}
\vskip -5pt
where $y$ denotes the output of the MoE layer, $x$ denotes the hidden representation from the previous layer, $\Psi$ denotes the selected expert set. We use top-1 routing, so $\Psi$ contains only the expert with the highest routing probability among the $N$ experts. $p_i(x+s)$ is the routing probability assigned to expert $i$, and $E_i(x)$ is the output of expert $i$.

To alleviate unbalanced expert assignment, we incorporate the load balancing loss~\cite{fedus2022switch} as an auxiliary loss $\mathcal{L}_{aux}$. The loss is scaled by $\alpha$, as shown in Eq.~\ref{eq:auxLoss}. Following Switch Transformer, we use tokens to denote the routing units in the MoE layer. In our DP, these tokens correspond to phoneme-level hidden states. $f_i$ in Eq.~\ref{eq:tokenFrac} measures the fraction of tokens assigned to expert $i$ within a mini-batch $\mathcal{B}$, while $P_i$ in Eq.~\ref{eq:probFrac} denotes the average routing probability of expert $i$. $T$ is the total number of tokens in $\mathcal{B}$.
\vskip -15pt
\begin{align}
	\mathcal{L}_{aux} = \alpha\cdot N\cdot \sum_{i=1}^N f_{i} \cdot P_{i}
	\label{eq:auxLoss}
\end{align}
\vskip -15pt
\begin{align}
f_{i} = \frac{1}{T}\sum_{token \in \mathcal{B}}\mathds{1}\{\arg\max_{j=1,\ldots,N}\ p_j(token) = i\}
\label{eq:tokenFrac}
\end{align}
\vskip -15pt
\begin{align}
	P_{i} = \frac{1}{T}\sum_{token \in \mathcal{B}}p_{i}(token)
	\label{eq:probFrac}
\end{align}

Finally, the overall optimization objective for the MoE-DP is defined as:
\vskip -15pt
\begin{align}
	\mathcal{L}_{dur} = \mathcal{L}_{mas} + \mathcal{L}_{aux}.
	\label{eq:moeLoss}
\end{align}

\subsection{Vocoder and Discriminators}
\label{sec:vocoder}

To render the diverse duration patterns modeled by MoE-DP with both high waveform quality and low inference cost, we combine a VOCOS-style generator with CoMBD and SBD, as shown in Fig.~\ref{fig:Overall}. The VOCOS-style generator provides efficient frequency-domain synthesis, while CoMBD and SBD strengthen waveform supervision across temporal resolutions and frequency bands.

We employ a VOCOS-style generator based on ConvNeXt blocks~\cite{siuzdak2023vocos,liu2022convnet}. Unlike HiFi-GAN-based generators that progressively upsample acoustic representations in the time domain, this design directly predicts complex STFT coefficients and reconstructs the waveform through ISTFT. To integrate it into the VITS end-to-end training pipeline, we condition the generator on the posterior latent variable $z$ and speaker embedding $s$:
\vskip -10pt
\begin{align}
    \hat{w} = \mathrm{ISTFT}(\mathrm{Backbone}(z + s)),
    \label{eq:vocos}
\end{align}
where $\hat{w}$ denotes the reconstructed waveform. This frequency-domain formulation avoids stacked temporal upsampling and supports efficient waveform generation.

For adversarial training, we adopt the Collaborative Multi-Band Discriminator (CoMBD) and Sub-Band Discriminator (SBD)~\cite{bak2023avocodo}.

\textbf{Collaborative Multi-Band Discriminator (CoMBD)}: This discriminator operates on waveform signals at different resolutions. To reduce computational cost, each waveform resolution is processed using the same multi-scale discriminator (MSD)~\cite{kumar2019melgan}. This design evaluates global waveform coherence and local waveform structures across resolutions, providing multi-scale supervision for temporal continuity.

\textbf{Sub-Band Discriminator (SBD)}: This discriminator employs PQMF analysis~\cite{nguyen1994near} to decompose the speech waveform into multiple sub-band signals, enabling discrimination across different frequency ranges. Multi-scale dilated convolutions further extract sub-band features at different receptive fields, providing frequency-dependent supervision for spectral consistency.

\subsection{Model and Training Configuration}

For the Text Encoder, Speaker Encoder, Posterior Encoder, and Flow, we adopt the original VITS hyperparameters. For the MoE-DP, the convolution part consists of two 1D convolution blocks with a kernel size of 3 and a hidden dimension of 192. The MoE part contains two Switch-Transformer blocks, each with 8 experts, 4 attention heads, and a hidden dimension of 192. We use top-1 routing with a capacity factor of 1.0. For the vocoder, the hop length and FFT size are set to 256 and 1024, respectively. It contains 8 ConvNeXt blocks, each with an intermediate dimension of 1536 and a hidden dimension of 512.

Our system is trained using the AdamW optimizer with $\beta_1=0.8$ and $\beta_2=0.99$. The initial learning rate is $2\times10^{-4}$, with a decay factor of 0.999 applied after each epoch. The batch size $\mathcal{B}$ is 24, and the model is trained on 4 NVIDIA RTX 3090 GPUs.

\section{Experimental Setup}
\label{sec:evaluation}

\subsection{Overall Quality Assessment}

We evaluate overall synthesis quality on LJSpeech~\cite{ljspeech17} and VCTK~\cite{yamagishi2019vctk}, two widely used benchmarks for single-speaker and multi-speaker TTS, using 1,324 and 1,500 held-out samples, respectively. All systems evaluated on the same dataset synthesize the same test sentences. In our experiments, LJSpeech and VCTK are used at sampling rates of 22.05~kHz and 16~kHz, respectively.

We use Mean Opinion Score (MOS) and Word Error Rate (WER) to evaluate synthesis quality and additionally report model and discriminator sizes. Perceptual quality is assessed using a standard five-point MOS scale~\cite{itu1996p800}. The same ten annotators evaluated all systems. Each sample received 10 ratings, and mean MOS is reported with 95\% confidence intervals. WER measures content accuracy and intelligibility using Whisper-large-v3~\cite{radford2022whisper}, with the same transcription and calculation pipeline applied to all systems. Although different systems may generate audio at different sampling rates, all outputs are resampled to 16~kHz using the same procedure before MOS and WER evaluation. Model and discriminator sizes are calculated directly from the evaluated checkpoints.

For external baseline comparison, we consider two categories of systems: dataset-specific TTS systems, represented by FastSpeech 2~\cite{ren2020fastspeech} and StyleTTS 2~\cite{li2023styletts}, and large-scale reference-conditioned zero-shot systems, represented by F5-TTS~\cite{chen2024f5} and Spark-TTS~\cite{wang2025spark}. The former are evaluated using their released dataset-specific checkpoints (no official VCTK dataset model, so VCTK results are unavailable), whereas the latter use their official pretrained checkpoints with speaker-matched reference utterances different from the target utterances. All external baselines retain their original training configurations and are evaluated using their complete released inference pipelines, without retraining or architectural modification. For internal ablations, VITS Origin and all FNH-TTS variants are trained and evaluated by us using the same data splits, training steps, random seeds, and evaluation protocol, differing only in the ablated components.

\subsection{Prosody Modeling Assessment}

To evaluate duration-related prosody modeling, we use LibriTTS~\cite{zen2019libritts} (sampling rate: 24~kHz). The train-clean-100 and train-clean-360 subsets, collectively denoted as Libri460, are used to train our VITS-based systems, while test-clean is used for evaluation. We select LibriTTS because phoneme-duration annotations are provided by Ji et al.~\cite{ji2024textrolspeech}, whereas LJSpeech and VCTK do not provide corresponding annotations.

To compute duration-category accuracy, we follow the evaluation script provided by ControlSpeech~\cite{ji2025controlspeech}. Montreal Forced Alignment (MFA)~\cite{mcauliffe2017montreal} first aligns each synthesized utterance with its phoneme sequence to determine the phoneme boundaries, from which the average phoneme duration is calculated. An utterance is classified as fast if its average phoneme duration is below 0.0727\,s, normal if it falls between 0.0727\,s and 0.1107\,s, and slow if it is above 0.1107\,s. Duration-category accuracy is computed by comparing the synthesized category with its manually annotated reference label. We also report MOS and WER following the overall quality assessment protocol and visualize utterance-level duration distributions on LJSpeech and VCTK to examine natural duration variation and speaker-dependent speaking-rate patterns.

The external comparison systems and inference protocols follow the overall quality assessment, with the FastSpeech 2 checkpoint provided by ESPnet. For internal comparisons, we include VITS+DDP, VITS+SDP~\cite{kim2021conditional}, and VITS2~\cite{kong2023vits2}, which cover three principal duration-prediction paradigms within the VITS framework: deterministic regression, flow-based stochastic prediction, and adversarially trained stochastic prediction, respectively. This selection enables a controlled comparison of MoE-DP with established DP designs while keeping the remaining VITS framework unchanged. All VITS-based systems, including FNH-TTS, are trained by us under the same controlled setting.

\subsection{Vocoder Reconstruction and Inference Efficiency}

We evaluate vocoder reconstruction and inference efficiency on the same LibriTTS test-clean subset used in the prosody assessment. Both reference and reconstructed audio are evaluated at the native sampling rate of 24~kHz, providing objective evidence complementary to the end-to-end MOS evaluation.

Because end-to-end synthesized speech and its reference often differ in length, directly applying reference-based metrics may conflate waveform reconstruction with duration and alignment errors. We therefore adopt an analysis-by-synthesis protocol that encodes each reference linear spectrogram into the posterior latent variable $z$ and reconstructs the waveform from $z$, thereby bypassing duration prediction and text--audio alignment. We report Multi-Scale STFT Loss (M-STFT)~\cite{yamamoto2020parallel}, Perceptual Evaluation of Speech Quality (PESQ)~\cite{rix2001perceptual}, Mel Cepstral Distortion (MCD)~\cite{kubichek1993mel}, Periodicity, and Voiced/Unvoiced F1 Score (V/UV F1)~\cite{morrison2021chunked}. RTF is measured with a batch size of 1 on an Intel Xeon Gold 5218R CPU and a single NVIDIA RTX 3090 GPU.

We compare the FNH-TTS vocoder configuration with the VITS vocoders based on HiFi-GAN~\cite{kong2020hifi} and HiFi-GANv2~\cite{su2021hifi}. All configurations are trained on the same Libri460 data and evaluated on test-clean under the same protocol, differing only in their vocoder-related configurations. The comparison is limited to VITS-based systems because the reconstruction protocol requires the posterior latent representation.

\section{Results}

This section evaluates FNH-TTS from three aspects: overall synthesis quality, duration-related prosody modeling, and vocoder reconstruction and inference efficiency.

\subsection{Synthesis Quality Comparison}
\label{sec:overallResults}

\begin{table*}[htbp]
\small
\centering
\caption{Overall synthesis quality and leave-one-component-out ablations. MOS is reported with 95\% confidence intervals. In each ablation, the removed component is replaced by its corresponding VITS component.}
%\vskip -5pt
\label{tab:finalResult}
\resizebox{\textwidth}{!}{
\begin{tabular}{lcccccc}
\hline
\multirow{2}{*}{\textbf{System}} &
\multicolumn{2}{c}{\textbf{LJSpeech}} &
\multicolumn{2}{c}{\textbf{VCTK}} &
\multirow{2}{*}{\textbf{\begin{tabular}[c]{@{}c@{}}Model\\Params\end{tabular}}} &
\multirow{2}{*}{\textbf{\begin{tabular}[c]{@{}c@{}}Disc.\\Params\end{tabular}}} \\
& \textbf{MOS} ($\uparrow$) & \textbf{WER} ($\downarrow$)
& \textbf{MOS} ($\uparrow$) & \textbf{WER} ($\downarrow$)
& & \\ \hline
\multicolumn{7}{l}{\textit{External baselines}} \\
FastSpeech 2 & $4.12 \pm 0.09$ & 7.02\% & -- & -- & 34.64M & 42.53M \\
StyleTTS 2   & $4.35 \pm 0.10$ & 5.78\% & -- & -- & 145.53M & 41.38M \\
F5-TTS       & $3.87 \pm 0.13$ & 8.70\% & $4.49 \pm 0.08$ & \textbf{2.24\%} & 337.09M & -- \\
Spark-TTS    & $4.10 \pm 0.11$ & 7.36\% & $4.38 \pm 0.12$ & 4.98\% & 506.63M & -- \\ \hline
\multicolumn{7}{l}{\textit{Controlled VITS systems}} \\
VITS Origin                  & $4.26 \pm 0.08$ & 3.41\% & $4.34 \pm 0.09$ & 4.11\% & 39.53M & 46.75M \\
\textbf{FNH-TTS}             & $\boldsymbol{4.48 \pm 0.10}$ & 2.59\% & $\boldsymbol{4.63 \pm 0.09}$ & 3.88\% & 47.73M & 27.07M \\
\quad w/o MoE-DP             & $4.44 \pm 0.09$ & 2.60\% & $4.27 \pm 0.09$ & 4.96\% & 40.89M & 27.07M \\
\quad w/o VOCOS              & $4.20 \pm 0.11$ & \textbf{2.42\%} & $4.43 \pm 0.11$ & 3.60\% & 46.37M & 27.07M \\
\quad w/o CoMBD/SBD          & $3.75 \pm 0.12$ & 9.74\% & $3.95 \pm 0.12$ & 10.58\% & 47.73M & 46.75M \\ \hline
\end{tabular}
}
\vskip -15pt
\end{table*}

Table~\ref{tab:finalResult} compares FNH-TTS with external TTS systems and controlled VITS-based variants. FNH-TTS achieves the highest mean MOS on both LJSpeech and VCTK and obtains higher mean MOS than VITS Origin, while its 47.73M model parameters remain substantially below those of the large-scale F5-TTS and Spark-TTS systems.

The leave-one-component-out ablations reveal the contribution of each component within the complete system. \textbf{(1) Removing MoE-DP.} MoE-DP contributes positively to FNH-TTS, with its effect most evident in the multi-speaker setting. Without MoE-DP, the VCTK mean MOS decreases from 4.63 to 4.27 and WER increases from 3.88\% to 4.96\%, both becoming worse than VITS Origin. \textbf{(2) Removing VOCOS.} VOCOS contributes to perceived synthesis quality, although this contribution is not reflected by WER. Removing it lowers mean MOS on both datasets while improving WER, illustrating that MOS and WER capture different aspects of synthesized speech. \textbf{(3) Removing CoMBD and SBD.} CoMBD and SBD provide the strongest support for final synthesis quality within the complete configuration. Their removal causes the largest and most consistent degradation, reducing mean MOS by approximately 0.7 and increasing WER to around 10\% on both datasets.

Based on the mean-MOS degradation caused by component removal, the ablations suggest an apparent contribution pattern of CoMBD/SBD $>$ MoE-DP $\approx$ VOCOS. However, this metric-level ordering describes only the sensitivity of the complete system to component removal and does not reveal the distinct modeling roles of these components or their underlying interaction. We therefore examine duration-related prosody modeling and vocoder reconstruction in Sections~\ref{sec:DPAnalysis} and~\ref{sec:vocoderAnalysis}, respectively, before further analyzing their interaction in Section~\ref{sec:dis}.

\subsection{Prosody Modeling Analysis}
\label{sec:DPAnalysis}

\begin{table}[htbp]
\small
\centering
\caption{Duration-related prosody evaluation on LibriTTS test-clean. MOS is reported with 95\% confidence intervals.}
\vskip -5pt
\label{tab:durComp}
\setlength{\tabcolsep}{4.5pt}
\renewcommand{\arraystretch}{1.05}
\begin{tabular}{lccc}
\hline
\textbf{System} &
\textbf{Dur. Acc.} ($\uparrow$) &
\textbf{MOS} ($\uparrow$) &
\textbf{WER} ($\downarrow$) \\ \hline
\multicolumn{4}{l}{\textit{External baselines}} \\
FastSpeech 2 & 59.80\% & $4.37 \pm 0.06$ & 2.12\% \\
StyleTTS 2   & 59.28\% & $4.41 \pm 0.08$ & \textbf{1.58\%} \\
F5-TTS       & 11.17\% & $4.35 \pm 0.10$ & 4.21\% \\
Spark-TTS    & 66.77\% & $4.36 \pm 0.09$ & 5.12\% \\ \hline
\multicolumn{4}{l}{\textit{VITS-based duration baselines}} \\
VITS+DDP & 61.26\% & $4.37 \pm 0.08$ & 3.42\% \\
VITS+SDP & 65.76\% & $4.27 \pm 0.09$ & 6.06\% \\
VITS2    & 60.36\% & $4.33 \pm 0.08$ & 3.34\% \\ \hline
\textbf{FNH-TTS} &
\textbf{67.07\%} &
$\boldsymbol{4.42 \pm 0.07}$ &
3.15\% \\ \hline
\end{tabular}
\vskip -15pt
\end{table}

Using the annotated duration categories in LibriTTS test-clean, Table~\ref{tab:durComp} directly evaluates duration-related prosody together with perceptual quality MOS and intelligibility WER. FNH-TTS achieves the highest duration-category accuracy of 67.07\% and the highest mean MOS of 4.42. These point estimates are the highest among both the external systems and the three representative VITS-based duration-modeling baselines.

The metric rankings are again different: StyleTTS 2 obtains the lowest WER, whereas FNH-TTS achieves higher duration-category accuracy and mean MOS. This result reinforces that WER and MOS reflect different aspects of synthesized audio. Because TTS naturalness is ultimately judged by listeners, WER should be interpreted together with MOS and duration-specific metrics rather than as a substitute for perceptual evaluation.

\begin{figure}[htbp]
%\vskip -15pt
  \centering
  \includegraphics[width=\linewidth]{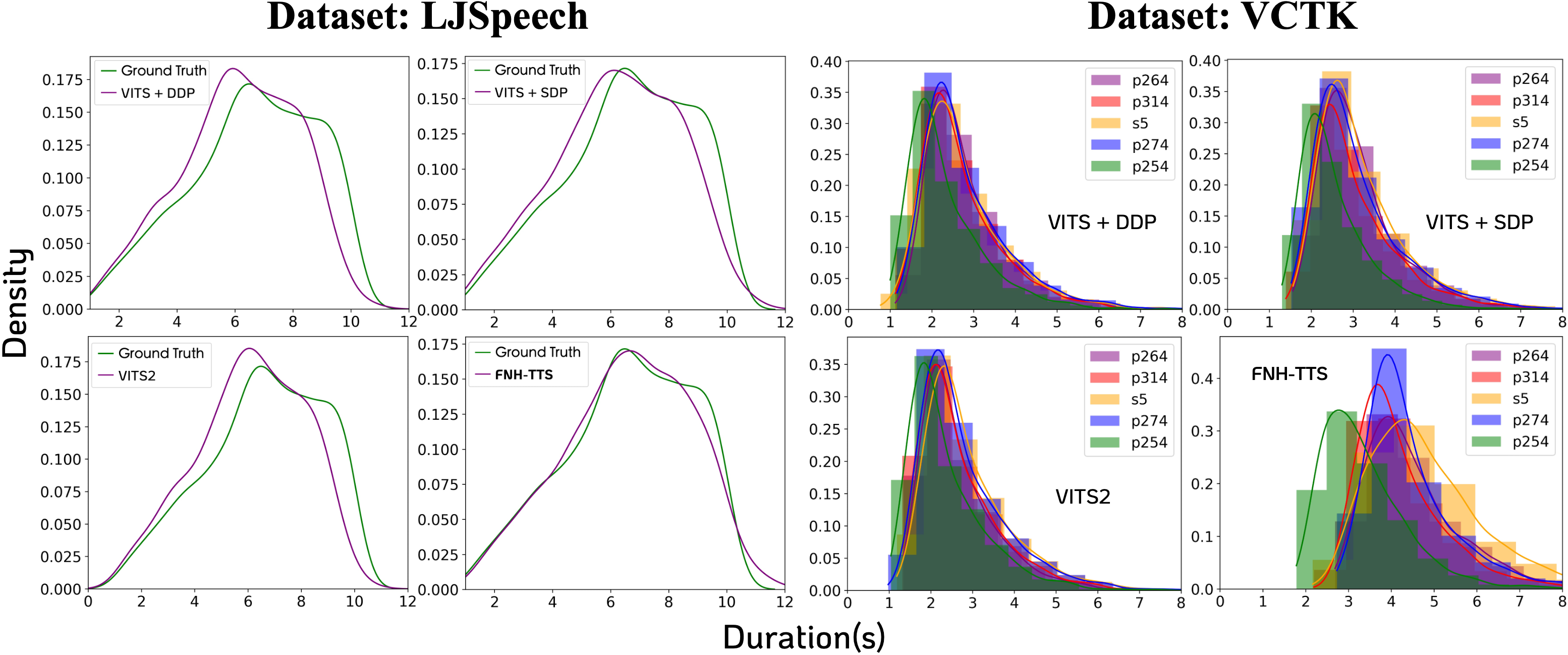}
  \vskip -10pt
  \caption{Utterance-level duration distributions of different systems on LJSpeech and VCTK.}
  \label{fig:DurCompare}
  \vskip -15pt
\end{figure}

To complement the duration-category evaluation, we visualize the utterance-level duration distributions of synthesized speech in Fig.~\ref{fig:DurCompare}. On LJSpeech, FNH-TTS more closely follows the overall shape of the reference distribution, whereas VITS+DDP and VITS2 exhibit sharper peaks and larger distributional deviations. On VCTK, FNH-TTS produces more distinct speaker-dependent duration distributions, while the other VITS-based systems show greater overlap among speakers.

Together, the duration-category results and distribution visualizations support that FNH-TTS better captures duration-related prosody. However, the current evidence has three limitations: \textbf{(1)} duration-category accuracy does not quantify fine-grained phoneme-duration distribution differences; \textbf{(2)} this labeled metric is available only for LibriTTS; and \textbf{(3)} complete-system comparisons do not isolate the contribution of MoE-DP. Therefore, Section~\ref{sec:roleMOE} presents an additional controlled analysis on LJSpeech and VCTK by comparing the predicted phoneme-duration sequences of each variant with those of the complete FNH-TTS system. Because this analysis uses the FNH-TTS output as an empirical anchor rather than manually annotated ground truth, it measures component-level consistency with the duration behavior of FNH-TTS instead of absolute duration accuracy. We therefore treat it as complementary diagnostic evidence and present it in the Discussion rather than as a primary evaluation result.

\subsection{Vocoder Performance Evaluation}
\label{sec:vocoderAnalysis}

\begin{table}[htbp]
\centering
\caption{Vocoder reconstruction quality and inference efficiency on LibriTTS test-clean under the analysis-by-synthesis setting. Bold and underlined values denote the best and second-best results, respectively.}
\label{tab:vocoderComp}
\begingroup
\small
\renewcommand{\arraystretch}{1.08}

\begin{tabular*}{\linewidth}{@{\extracolsep{\fill}}lccccc@{}}
\multicolumn{6}{l}{\textbf{(a) Reconstruction quality}}\\[-1pt]
\toprule
\textbf{System} &
\textbf{M-STFT($\downarrow$)} &
\textbf{PESQ($\uparrow$)} &
\textbf{MCD($\downarrow$)} &
\textbf{Periodicity($\downarrow$)} &
\textbf{V/UV F1($\uparrow$)} \\
\midrule
VITS (HiFi-GAN) &
\underline{1.208} &
\textbf{2.441} &
\underline{1.677} &
\textbf{0.160} &
\textbf{0.928} \\
VITS (HiFi-GANv2) &
1.246 &
2.231 &
1.814 &
0.166 &
0.922 \\
\textbf{FNH-TTS} &
\textbf{1.207} &
\underline{2.425} &
\textbf{1.654} &
\underline{0.162} &
\underline{0.927} \\
\bottomrule
\end{tabular*}

\vspace{5pt}

\begin{tabular*}{\linewidth}{@{\extracolsep{\fill}}lcc@{}}
\multicolumn{3}{l}{\textbf{(b) Inference efficiency}}\\[-1pt]
\toprule
\textbf{System} &
\textbf{CPU RTF($\downarrow$)} &
\textbf{GPU RTF($\downarrow$)} \\
\midrule
VITS (HiFi-GAN) &
0.352 &
0.00748 \\
VITS (HiFi-GANv2) &
\underline{0.181} &
\underline{0.00659} \\
\textbf{FNH-TTS} &
\textbf{0.046} &
\textbf{0.00460} \\
\bottomrule
\end{tabular*}

\endgroup
\end{table}

Table~\ref{tab:vocoderComp} evaluates complementary aspects of vocoder performance: M-STFT and MCD reflect spectral reconstruction, PESQ estimates perceptual quality, and Periodicity and V/UV F1 measure periodic structure and voicing consistency.

The FNH-TTS vocoder configuration achieves the lowest M-STFT and MCD, demonstrating its advantage in spectral reconstruction. Its PESQ, Periodicity, and V/UV F1 are numerically close to the best HiFi-GAN results, and it outperforms HiFi-GANv2 across all five reconstruction metrics. These results show that the proposed configuration provides a strong overall balance between spectral fidelity, perceptual quality, and voicing consistency, although it does not rank first on every metric.

FNH-TTS also provides a substantial efficiency advantage. Compared with the original VITS vocoder, it achieves approximately $7.7\times$ faster CPU inference and $1.6\times$ faster GPU inference. The reconstruction results characterize the complete FNH-TTS vocoder configuration, whereas the RTF improvement mainly reflects the VOCOS-style generator because CoMBD and SBD are used only during training.

Together with the leave-one-component-out results in Table~\ref{tab:finalResult}, where removing VOCOS lowers the mean MOS on both datasets, these findings support its contribution to perceived synthesis quality and waveform-generation efficiency. Overall, the VOCOS-style generator offers a favorable quality--efficiency trade-off within FNH-TTS rather than improving efficiency at the expense of reconstruction quality.

\section{Discussion}
\label{sec:dis}

The preceding results show that FNH-TTS achieves strong overall and duration-related performance, while its components contribute differently to synthesis quality. However, the leave-one-component-out results alone do not fully explain the roles of the individual components. We therefore conduct additional controlled analyses to isolate the contribution of MoE-DP to duration modeling and examine how the vocoder-side components complement it during waveform generation.

\subsection{The Role of MoE-DP in Duration Modeling}
\label{sec:roleMOE}

The labeled LibriTTS results show that the complete FNH-TTS system achieves better duration-related prosody modeling, but the complete-system comparison does not isolate the contribution of MoE-DP. We therefore analyze whether this duration behavior primarily comes from MoE-DP and remains stable across vocoder-side configurations at both the utterance and phoneme levels.

\begin{figure}[htbp]
  \centering
  \includegraphics[width=\linewidth]{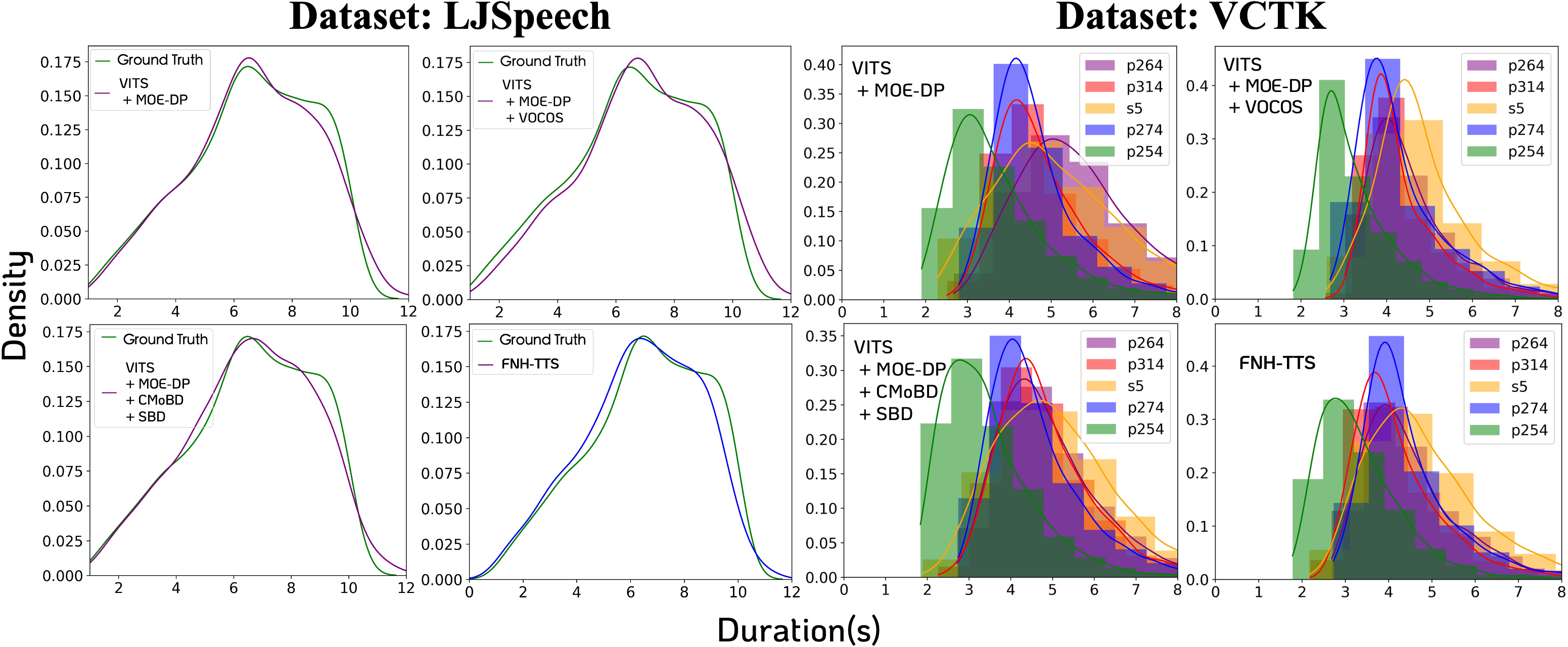}
  \caption{Utterance-duration distributions of MoE-DP-based systems under different vocoder-side configurations on LJSpeech and VCTK.}
  \label{fig:DurAblation}
\end{figure}

Figure~\ref{fig:DurAblation} shows that \textbf{(1)} all MoE-DP-based configurations produce similar utterance-duration distributions that remain close to the reference distribution on LJSpeech; \textbf{(2)} they preserve visible speaker-dependent duration differences on VCTK; and \textbf{(3)} together with the non-MoE comparisons in Fig.~\ref{fig:DurCompare}, these results suggest that the main duration characteristics of FNH-TTS are associated with MoE-DP and are not removed by VOCOS or CoMBD/SBD.

To further quantify the contribution of MoE-DP, we use the predicted phoneme-duration sequence of the complete FNH-TTS system as an anchor distribution and compute its Jensen--Shannon (JS) divergence from the DP output of each variant. This choice is motivated by the highest duration-category accuracy achieved by FNH-TTS on the annotated LibriTTS benchmark. For the same input, all controlled variants use the same phoneme sequence, enabling direct comparison between their DP outputs; a lower JS value indicates duration behavior closer to FNH-TTS. MFA is not used because its phoneme inventory does not match that of our model, preventing one-to-one comparison with the DP output. We report the mean and variance over all test utterances and omit FNH-TTS because its self-divergence is zero.

\begin{table}[htbp]
\small
\centering
\caption{Jensen--Shannon divergence between the predicted phoneme-duration sequences of each system and those of the complete FNH-TTS system.}
\label{tab:JSDivergence}
\begin{tabular*}{\linewidth}{@{\extracolsep{\fill}}lcccc@{}}
\toprule
\multirow{2}{*}{\textbf{System}} &
\multicolumn{2}{c}{\textbf{LJSpeech}} &
\multicolumn{2}{c}{\textbf{VCTK}} \\
\cmidrule(lr){2-3}\cmidrule(lr){4-5}
& \textbf{Mean} & \textbf{Var.} & \textbf{Mean} & \textbf{Var.} \\
\midrule
VITS + DDP                        & 0.057 & 0.0002 & 0.044 & 0.0002 \\
VITS + SDP                        & 0.087 & 0.0003 & 0.066 & 0.0001 \\
\midrule
VITS + MoE-DP                     & 0.053 & 0.0002 & 0.039 & 0.0001 \\
VITS + MoE-DP + VOCOS             & 0.053 & 0.0002 & 0.043 & 0.0001 \\
VITS + MoE-DP + CoMBD/SBD         & 0.047 & 0.0002 & 0.041 & 0.0004 \\
\bottomrule
\end{tabular*}
\end{table}

Table~\ref{tab:JSDivergence} shows that \textbf{(1)} all MoE-DP-based variants obtain lower mean JS divergence than VITS+DDP and VITS+SDP on both datasets, indicating that the main duration behavior of FNH-TTS follows MoE-DP; \textbf{(2)} the divergence remains within the lower range of the MoE-DP variants after introducing VOCOS or CoMBD/SBD, showing that this behavior is preserved across vocoder-side configurations; and \textbf{(3)} because the analysis measures similarity to FNH-TTS rather than absolute accuracy against manually annotated durations, it is treated only as complementary component-attribution evidence. Together with the labeled LibriTTS evaluation, these results support MoE-DP as the primary source of the duration-modeling characteristics of FNH-TTS, while the effects of the vocoder-side components on final waveform synthesis are examined next.

\subsection{Interaction Between Duration Modeling and Vocoder-Side Components}
\label{sec:roleVocoder}

The quantitative results identify three complementary roles in FNH-TTS. \textbf{(1)} MoE-DP is the primary source of the duration-modeling characteristics, as shown in Section~\ref{sec:roleMOE}. \textbf{(2)} CoMBD/SBD provide the strongest support for final synthesis quality, as removing them causes the largest degradation in MOS and WER in Table~\ref{tab:finalResult}. \textbf{(3)} VOCOS contributes to perceived quality and waveform-generation efficiency. Removing it lowers the mean MOS on both datasets, although this change is not reflected by WER; Table~\ref{tab:vocoderComp} further shows that the complete vocoder configuration combines strong reconstruction quality with the lowest RTF. Since CoMBD/SBD are used only during training, the efficiency gain mainly reflects the VOCOS-style generator.

\begin{figure*}[htbp]
  \centering
  \includegraphics[width=\linewidth]{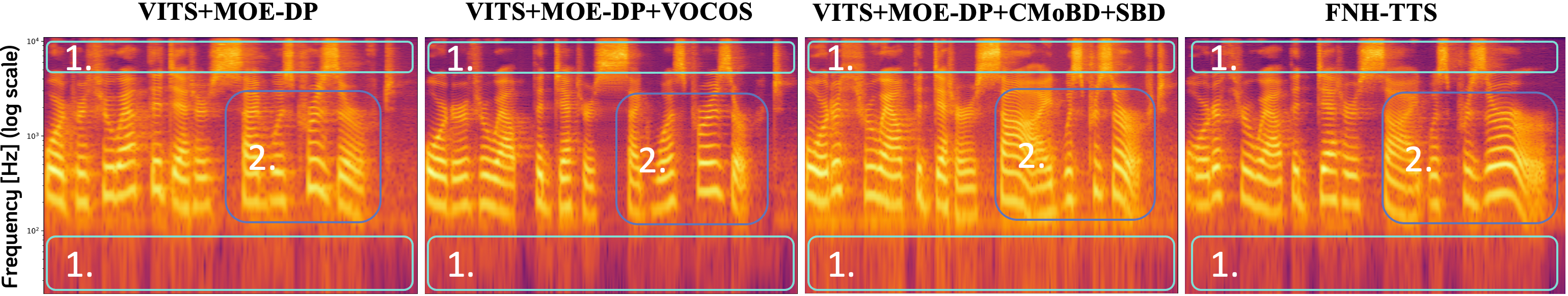}
  \caption{Spectrograms of an LJSpeech sample synthesized by VITS + MoE-DP, VITS + MoE-DP + VOCOS, VITS + MoE-DP + CoMBD/SBD, and the complete FNH-TTS system.}
  \label{fig:spectrograms}
\end{figure*}

To illustrate how these components affect waveform synthesis, Fig.~\ref{fig:spectrograms} compares spectrograms produced by four controlled configurations.

The selected example illustrates three differences. \textbf{(1)} VITS + MoE-DP exhibits weaker spectral details and less continuous time--frequency structures in the highlighted high- and low-frequency regions. \textbf{(2)} Adding VOCOS produces sharper spectral textures, while adding CoMBD/SBD results in smoother temporal transitions. \textbf{(3)} The complete FNH-TTS system combines clearer spectral details with more coherent temporal structures. Because this comparison is based on a selected example, it should be interpreted as a qualitative illustration rather than general evidence of a causal relationship.

Taken together, the quantitative and qualitative results support complementary roles for MoE-DP and the vocoder-side components under the evaluated configurations. They do not establish that richer duration variation itself causes the degradation of the original HiFi-GAN-based generator, nor do they exclude other training- or optimization-related explanations. Instead, the results motivate the joint design of FNH-TTS: MoE-DP primarily improves duration modeling, CoMBD/SBD support final synthesis quality, and VOCOS contributes to perceived quality and efficient waveform generation while the complete vocoder configuration maintains strong reconstruction quality.

\section{Conclusion}

In this paper, we propose FNH-TTS, a VITS-based NAR TTS system that jointly strengthens duration modeling and waveform generation. MoE-DP captures diverse contextual and speaker-dependent duration patterns, while the VOCOS-style generator enables efficient waveform generation and CoMBD/SBD provide complementary adversarial supervision.

Experiments on LJSpeech, VCTK, and LibriTTS demonstrate strong synthesis quality, improved duration-related performance, competitive waveform reconstruction, and substantially improved vocoder inference efficiency. The leave-one-component-out ablations and controlled analyses support distinct but complementary roles: MoE-DP is the primary source of the duration-modeling characteristics, CoMBD/SBD provide the strongest support for final synthesis quality, and VOCOS contributes to perceived quality and efficient waveform generation while the complete vocoder configuration maintains competitive reconstruction quality. These findings characterize the empirical roles of the proposed components without establishing a specific causal relationship between duration variation and vocoder behavior.

Overall, our results suggest that duration modeling and waveform generation should be considered jointly in end-to-end TTS design. Future work will further investigate their interaction and develop more reliable fine-grained prosody evaluation methods that complement perceptual and intelligibility-oriented metrics such as MOS and WER.

 \bibliographystyle{splncs03_unsrt}
 \bibliography{mybibliography}
%Author, F.: Article title. Journal \textbf{2}(5), 99--110 (2016)
%Author, F., Author, S.: Title of a proceedings paper. In: Editor,
%F., Editor, S. (eds.) CONFERENCE 2016, LNCS, vol. 9999, pp. 1--13.
%Springer, Heidelberg (2016). \doi{10.10007/1234567890}
%Author, F., Author, S., Author, T.: Book title. 2nd edn. Publisher,
%Author, A.-B.: Contribution title. In: 9th International Proceedings
%on Proceedings, pp. 1--2. Publisher, Location (2010)
%LNCS Homepage, \url{http://www.springer.com/lncs}, last accessed 2023/10/25
%\end{thebibliography}
\end{document}